\documentclass[aps,floatfix,showpacs,preprint,preprintnumbers,superscriptaddress,amsmath,amssymb,prb]{revtex4-1}

\usepackage{graphicx}% Include figure files

\begin{document}

\title{\boldmath Magnetic transition, long-range order, and moment fluctuations \\ in the pyrochlore iridate Eu$_2$Ir$_2$O$_7$}

\author{Songrui Zhao}
\altaffiliation{Present address: Department of Electrical and Computer Engineering, McGill University, Montreal, Quebec, Canada H3A 2A7.}
\affiliation{Department of Physics and Astronomy, University of California, Riverside, California 92521, USA}
\author{J.~M. Mackie}
\affiliation{Department of Physics and Astronomy, University of California, Riverside, California 92521, USA}
\author{D.~E. MacLaughlin}
\affiliation{Department of Physics and Astronomy, University of California, Riverside, California 92521, USA}
\author{O.~O. Bernal}
\affiliation{Department of Physics and Astronomy, California State University, Los Angeles, California 90032, USA}
\author{J.~J. Ishikawa}
\affiliation{Institute for Solid State Physics, University of Tokyo, Kashiwa 277-8581, Japan}
\author{Y. Ohta}
\affiliation{Institute for Solid State Physics, University of Tokyo, Kashiwa 277-8581, Japan}
\author{S. Nakatsuji}
\affiliation{Institute for Solid State Physics, University of Tokyo, Kashiwa 277-8581, Japan}

\date{\today}

\begin{abstract}
Muon spin rotation and relaxation experiments in the pyrochlore iridate~Eu$_2$Ir$_2$O$_7$ yield a well-defined muon spin precession frequency below the metal-insulator/antiferromagnetic transition temperature~$T_M = 120$~K, indicative of long-range commensurate magnetic order and thus ruling out quantum spin liquid and spin-glass-like ground states. The dynamic muon spin relaxation rate is temperature-independent between 2~K and ${\sim}T_M$ and yields an anomalously long Ir$^{4+}$ spin correlation time, suggesting a singular density of low-lying spin excitations. Similar behavior is found in other pyrochlores and geometrically frustrated systems, but also in the unfrustrated iridate BaIrO$_3$. Eu$_2$Ir$_2$O$_7$ may be only weakly frustrated; if so, the singularity might be associated with the small-gap insulating state rather than frustration.
\end{abstract}


\maketitle

Geometrical frustration of collinear near-neighbor spin interactions is a consequence of the corner-shared tetrahedral structure of pyrochlore transition-metal oxides, and has motivated considerable study of these materials.\cite{*[{For a review, see }]GGG10} Compounds in the pyrochlore iridate family $R_2$Ir$_2$O$_7$, where $R$ is a trivalent lanthanide, are particularly interesting: Ir$^{4+}$ ($5d^5$) is expected to be a low-spin $S = 1/2$ ion, and the behavior of the Ir-derived conduction band is unusual.\cite{MWNY07} For $R = \mathrm{Pr}$, Nd, Sm, and Eu these compounds exhibit metallic behavior at high temperatures, while for $R = \mathrm{Gd}$, Tb, Dy, Ho, Er, Yb, and Y they are semiconducting. This crossover was attributed to reduction of the width of the Ir$^{4+}$-derived band as the $R$ ionic radius decreases across the rare-earth series.\cite{BlWh80,*KWK98} 

In early studies\cite{TWH01} spin-glass-like ordering was reported for $R = \mathrm{Y}$, Lu, Sm, and Eu on the basis of bifurcation of field-cooled (FC) and zero-field-cooled (ZFC) magnetizations and little or no specific heat anomaly at a transition temperature~$T_M$. $^{151}$Eu M\"ossbauer studies of Eu$_2$Ir$_2$O$_7$\cite{ChSl78,TWH01} found no long-range magnetic ordering down to 4.2~K\@. Subsequently, metal-insulator (MI) transitions at $T_M$ with small specific heat anomalies were reported\cite{MWNY07} for $R = \mathrm{Nd}$, Sm, and Eu, and an exotic chiral spin-liquid metallic ground state\cite{NMMT06,*MNOT10} was found in Pr$_2$Ir$_2$O$_7$. The MI transitions were attributed to Ir$^{4+}$ $5d$ electrons, with complex antiferromagnetic (AFM) ordering.

Y$^{3+}$ and Lu$^{3+}$ are nonmagnetic, as is Eu$^{3+}$ in the Hund's-rule ground state $J = 0$ ($L = S$),\cite{MWNY07,BlWh80} so that only Ir$^{4+}$ $5d$ electrons contribute to magnetism in these compounds.\cite{TWH01} Magnetic ordering of localized Ir$^{4+}$ ions has been observed in a number of insulating iridates outside the pyrochlore family\cite{CXAC02,*CLCD04,*CDCP07,*MSTN08} and is quite anomalous, because overlap of the large Ir$^{4+}$ wave functions should result in metallic conduction via Ir-derived bands. In the case of (unfrustrated) Sr$_2$IrO$_4$ a detailed treatment\cite{KJMK08} involving strong spin-orbit coupling leads to the possibility of a Mott transition, however, and suggests an effective angular momentum $J_{\mathrm{eff}}=1/2$. Alternatively, a Slater transition, as found in the pyrochlore~Cd$_2$Os$_2$O$_7$,\cite{MTGF01,*PMB02b} is suggested by the second-order nature of the transition. 

Thus Eu$_2$Ir$_2$O$_7$ is a potential example of a geometrically frustrated system with $\mathrm{``spin"} = 1/2$, and as such is of considerable fundamental interest.\cite{GGG10} This Rapid Communication reports results of muon spin rotation and relaxation ($\mu$SR) experiments\cite{YaDdR11} on a polycrystalline sample of this compound. A well-defined muon-spin precession frequency is observed below $T_M$, indicating a uniform internal field and thus ruling out significant disorder; the magnetic order is commensurate and long-ranged. The dynamic muon-spin relaxation rate~$\lambda_d$ reflects anomalously slow spin fluctuations and remains constant to low temperatures. We speculate that this behavior might not be due solely to geometrical frustration, but may signal new low-lying spin excitations associated with a small-gap insulating state. The data show no critical slowing down of magnetic fluctuations as $T \rightarrow T_M$ from above, suggesting a mean-field-like transition.

Polycrystalline samples of Eu$_2$Ir$_2$O$_7$ were fabricated using a solid-state reaction technique.\cite{MMNM07} dc magnetization data (not shown) are consistent with previous results.\cite{TWH01,IOMN10u} $\mu$SR experiments were carried out at the M20 beam line at TRIUMF, Vancouver, Canada, using standard time-differential $\mu$SR\@.\cite{YaDdR11} A weak (25-Oe) magnetic field was applied parallel to the initial muon polarization, to decouple\cite{HUIN79} muon spins from nuclear dipolar fields in the paramagnetic state. Data were taken in a $^4$He gas-flow cryostat over the temperature range~2--200~K.

\begin{figure}[ht]
\includegraphics[clip=,width=0.45\textwidth]{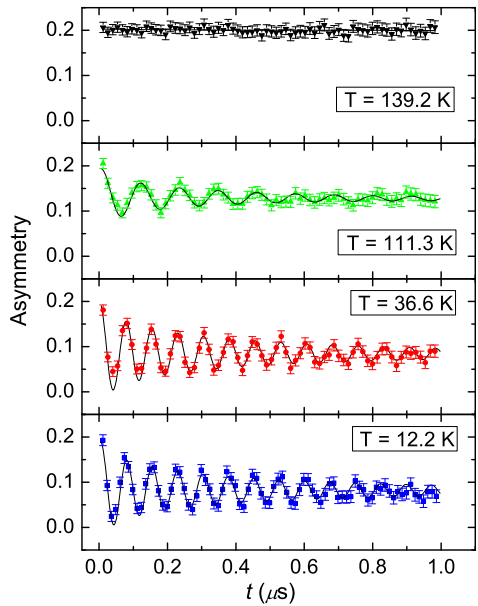}
\vspace{-15pt}
\caption{(Color online) Representative early-time asymmetry data~$A(t)$ in Eu$_2$Ir$_2$O$_7$, longitudinal field $= 25$~Oe. Solid curves: fits using Eq.~(\protect\ref{eq:fitfn}). }
\label{fig:asymm}
\end{figure}

Representative early-time asymmetry (signal amplitude) data $A(t)$ are shown in Fig.~\ref{fig:asymm}. 
Damped oscillations are observed below 120~K, due to precession of the muon spins in a quasistatic\cite{quasistatic} component~$\langle\mathbf{B}_\mathrm{loc}\rangle$ of the local field~$\mathbf{B}_\mathrm{loc}$ at muon sites. This confirms the magnetic transition found from the dc magnetization measurements. The oscillation is weakly damped except for the initial half cycle, indicating that $\langle\mathbf{B}_\mathrm{loc}\rangle$ is relatively homogeneous. The late-time asymmetry data (not shown) exhibit exponential relaxation, due solely to dynamic (thermal) fluctuations of $\mathbf{B}_\mathrm{loc}$.\cite{HUIN79} This relaxation is much slower than the oscillation damping rate, indicating that the latter reflects a quasistatic distribution of $\langle\mathbf{B}_\mathrm{loc}\rangle$.

The data were fit using the two-component asymmetry function
\begin{eqnarray}
A(t)& = & A_s\exp[-(\Lambda_st)^{K}]\cos(\omega_{\mu}t+\theta)\cr 
& & +A_d\exp(-\lambda_dt) \,.
\label{eq:fitfn}
\end{eqnarray}
The subscripts~$s$ and $d$ denote (quasi)static and dynamic components, respectively. The first term models the damped oscillation, with frequency~$\omega_\mu$ and spectrometer-dependent initial phase~$\theta$. Neither simple exponential damping nor a Bessel function (expected for an incommensurate spin density wave) gave good fits; the phenomenological stretched-exponential damping form of Eq.~(\ref{eq:fitfn}) was used instead, with relaxation rate~$\Lambda_s$ and stretching power~$K < 1$. The second term describes the late-time dynamic relaxation, which was well fit by a single exponential with rate~$\lambda_d$. 

The results of these fits are shown in Fig.~\ref{fig:asymm}. The data yield a single well-defined frequency (as does the Fourier transform, not shown), consistent with a commensurate magnetic structure and only one muon stopping site. The total initial asymmetry~$A(0) = A_s + A_d$ was found to be $\approx 0.21$ independent of temperature and applied field. 

\begin{figure}[ht]
\includegraphics[clip=,width=0.45\textwidth]{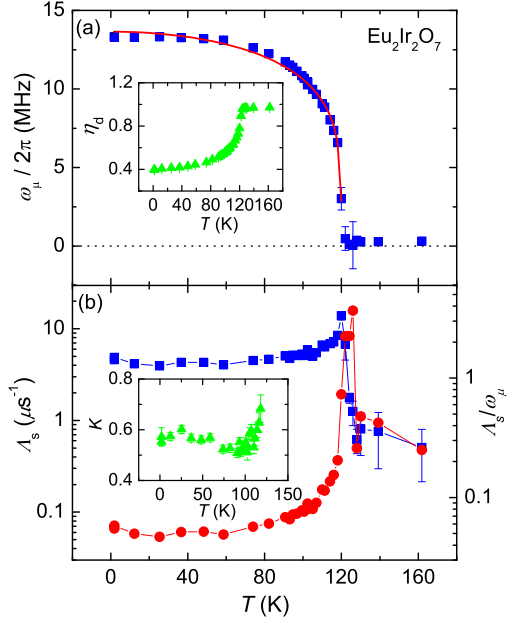}
%\vspace{-15pt}
\caption{(Color online) (a)~Temperature dependence of muon spin precession frequency~$\omega_{\mu}/2\pi$ in Eu$_2$Ir$_2$O$_7$. Inset: late-time fraction~$\eta_d$. The curve is a guide to the eye. (b)~Temperature dependence of quasistatic muon spin relaxation rate $\Lambda_s$ (squares, left axis) and fractional width of field distribution $\Lambda_s/\omega_{\mu}$ (circles, right axis). Inset: stretching power $K$. }
\label{fig:static}
\end{figure}

The temperature dependence of $\omega_{\mu}/2\pi$ and $\Lambda_s$ from the fits are shown in Figs.~\ref{fig:static}(a) and 2(b), respectively. 
The abrupt onset of $\omega_\mu$ and hence $\langle\mathbf{B}_{\mathrm{loc}}\rangle$ below 120~K indicates a magnetic transition at this temperature. At $T = 2$~K $\omega_{\mu}/2\pi = 13.32(3)$~MHz, corresponding to $\langle B_{\mathrm{loc}}\rangle = \omega_{\mu} / \gamma_{\mu} = 987(2)$~G\@. A rough estimate of the static Ir$^{4+}$ moment~$\mu_{\mathrm{Ir}}$ is given by equating this value to the internal field~$4\pi\mu_{\mathrm{Ir}}/v_{\mathrm{Ir}}$ of a uniform Ir$^{4+}$ magnetization, where $v_{\mathrm{Ir}}$ is the volume per Ir ion. This yields $\mu_{\mathrm{Ir}} \approx 1.1\mu_B$, of the order of the moment expected for $J_{\mathrm{eff}} = 1/2$.\cite{KJMK08} The estimate is very crude, however, because neither the Ir$^{4+}$ magnetic structure nor the muon stopping site is known.

As shown in the inset to Fig.~\ref{fig:static}(a), the late-time fraction~$\eta_d = A_d/(A_s+A_d)$ approaches 1 as $T \rightarrow T_M$ from below. This is due to the disappearance of $\langle\mathbf{B}_{\mathrm{loc}}\rangle$, and is consistent with the behavior of $\omega_{\mu}(T)$. At 2~K $\eta_d = 0.39(1)$, close to the value 1/3 expected from a randomly-oriented $\langle\mathbf{B}_{\mathrm{loc}}\rangle$.\cite{HUIN79} The increase of $\eta_d$ as $T \to T_M$ is smooth rather than abrupt, suggesting a distribution of transition temperatures in the sample. 

The temperature dependence of $\Lambda_s$ is given in Fig.~\ref{fig:static}(b).\footnote{The signal amplitude associated with the nonzero $\Lambda_s$ above $T_M$ is very small [$\eta_d \approx 1$, cf.\ insert to Fig.~\protect\ref{fig:static}(a)], and is either an instrumental artifact or due to a few percent of ordered second phase in the sample.} The cusp at ${\sim}T_M$ is probably an artifact of the distribution of $T_M$ noted above rather than a critical divergence, since as discussed below there is no sign of critical slowing down in the dynamic relaxation rate~$\lambda_d$. The fractional width~$\Lambda_s/\omega_{\mu}$ of the spontaneous field distribution, also plotted in Fig.~\ref{fig:static}(b), is small (0.05--0.07) at low temperatures and then increases rapidly as $T \to T_M$. Thus the local field is nearly uniform except in the neighborhood of $T_M$; this, like the behavior of $\eta_d$ noted above, suggests a distribution of $T_M$. 

The stretching power~$K$ for the quasistatic damping, shown in the inset of Fig.~\ref{fig:static}(b), parameterizes the shape of the distribution of $\langle B_{\mathrm{loc}}\rangle$: for small $K$ the wings of the distribution become more prominent.\cite{John06} The value of $K$ is temperature-independent ($\sim$0.55) at low temperatures and increases as $T \to T_M$. 

\begin{figure}[ht]
\includegraphics[clip=,width=0.45\textwidth]{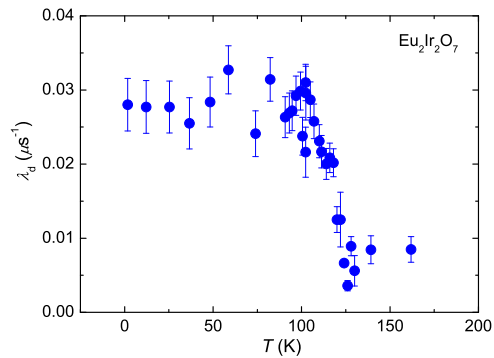}
\vspace{-15pt}
\caption{(Color online) Temperature dependence of the dynamic muon spin relaxation rate $\lambda_d$ in Eu$_2$Ir$_2$O$_7$. }
\label{fig:dynamic}
\end{figure}

The simple exponential form of the late-time relaxation data indicates that the dynamic muon spin relaxation, like $\langle{B}_{\mathrm{loc}}\rangle$ (but unlike $T_M$), is homogeneous. The temperature dependence of the dynamic relaxation rate~$\lambda_d$ is given in Fig.~\ref{fig:dynamic}. 
We note two features: (i)~$\lambda_d = 0.029(3)\ \mu\mathrm{s}^{-1}$ is constant below ${\sim}$100~K, and (ii)~with decreasing temperature there is an unusual step-like increase in $\lambda_d$ below $T_M$ but no sign of the paramagnetic-state divergence that is often found in frustrated and unfrustrated magnets\protect\cite{YDdRGM96,DKCG06,YDdRCM08,MNNH08} due to critical slowing down of dynamic fluctuations. This absence suggests a mean-field-like transition. 

The relation between dynamic muon relaxation and the moment fluctuations that cause it is generally complex. Limiting cases are (A)~quasistatic (slow) fluctuations of $\langle\mathbf{B}_\mathrm{loc}(t)\rangle$ with zero long-time average, where the relaxation time is essentially the correlation time of $\langle\mathbf{B}_\mathrm{loc}(t)\rangle$,\cite{HUIN79} and (B)~fluctuations~$\delta\mathbf{B}_\mathrm{loc}$ about a nonzero static $\langle\mathbf{B}_\mathrm{loc}\rangle$, i.e., $\mathbf{B}_\mathrm{loc}(t) = \langle\mathbf{B}_\mathrm{loc}\rangle + \delta\mathbf{B}_\mathrm{loc}(t)$;\cite{UYHS85} here the relaxation rate depends on the magnitude and stochastic properties of $\delta\mathbf{B}_\mathrm{loc}(t)$. Case~A describes dynamic relaxation in a paramagnet with extremely slow spin dynamics, and yields a fluctuation rate~$\sim \lambda_d \approx 3 \times 10^4~\mathrm{s}^{-1}$. The data cannot rule this scenario out in Eu$_2$Ir$_2$O$_7$ but it seems quite unlikely, given the phase-transition-like behavior of the muon spin precession frequency (Fig.~\ref{fig:static}) and the fact that a kilohertz fluctuation rate would be many orders of magnitude lower than any other frequency in the system. We therefore assume Case~B in further discussion of the dynamic relaxation.

In the motional narrowing limit~$\omega_{f}\tau_c \ll 1$ $\lambda_d \approx {\omega_{f}}^2 \tau_c$, where $\omega_{f} = \delta B_{\mathrm{loc}}/\gamma_\mu$ is the fluctuating field amplitude in fre\-quency units and $\tau_c$ is the correlation time of the fluctuations. Assuming a maximum $\omega_{f}$ of the order of the full quasistatic field in frequency units ($\omega_{f} \lesssim \omega_{\mu} \approx 8.5 \times 10^7~\mathrm{s}^{-1}$), this yields an upper bound~$\tau_c^{-1} \lesssim 2.5 \times~10^{11}~\mathrm{s}^{-1}$, or $\hbar/k_B\tau_c \lesssim 2$~K\@. In ordinary antiferromagnets $\hbar/k_B\tau_c$ is of the order of the N\'eel temperature~$T_N$ for $T \lesssim T_N$\@.\cite{Mori56} For Eu$_2$Ir$_2$O$_7$, with $T_N = T_M = 120$~K, $\tau_c$ is therefore at least two orders of magnitude longer than expected. 

The combination of a well-defined muon spin precession frequency (Fig.~\ref{eq:fitfn}), i.e., homogeneous magnetic order, and the persistence of $\lambda_d$ to low temperatures (Fig.~\ref{fig:dynamic}) is unexpected. In conventional ordered magnets nuclear or muon spin relaxation below the ordering temperature is due to thermal spin-wave excitations, and $\lambda_d$ decreases with decreasing temperature as the thermal population of such excitations decreases. Such a conventional scenario seems to be ruled out in Eu$_2$Ir$_2$O$_7$. 

Persistent low-temperature muon spin relaxation is observed in a number of geometrically frustrated systems.\cite{KMFS01,DKCG96,DKCG06,MOMN09,CFNG09} It indicates an enormously enhanced and possibly singular density of low-lying excitations, but is not well understood. In compounds containing non-Kramers rare-earth ions with nonmagnetic crystal-field ground states, fluctuations of hyperfine-enhanced nuclear magnetism can couple to muon spins and lead to persistent relaxation.\cite{SMAT07} This mechanism requires rare-earth ions with magnetic Hund's-rule ground states. A similar effect is associated with the low-lying Eu$^{3+}$ spin-orbit-split $J \ge 1$ multiplets; this, however, results in reduction rather than enhancement of Eu nuclear moments.\cite{Elli57,ShMa81} The persistent spin dynamics in Eu$_2$Ir$_2$O$_7$ must therefore be electronic in origin and associated with Ir$^{4+}$ magnetism. 

The relatively high transition temperature of Eu$_2$Ir$_2$O$_7$ suggests that the AFM exchange constant is not much larger than $T_M$, in which case Eu$_2$Ir$_2$O$_7$ is a weakly frustrated material.\cite{Rami94} Noting that the unfrustrated iridate~BaIrO$_3$ also exhibits persistent muon spin relaxation,\cite{BBLH05} we consider the possibility that frustration may not be the primary cause of persistent relaxation in Eu$_2$Ir$_2$O$_7$ and we look for another mechanism. 

In iridate compounds, frustrated or unfrustrated, the large Ir $5d$ wave functions are expected to weaken the on-site repulsion relative to the width of the $5d$ conduction band. If an AFM state associated with a metal-insulator transition is nevertheless retained (perhaps because of strong spin-orbit coupling\cite{KJMK08}) but the electrons are not well localized, the gap energy~$\Delta_g$ can be comparable to $k_BT_M$. The resistivity of single-crystal Eu$_2$Ir$_2$O$_7$ in fact yields a maximum gap value~$\approx 10~\mathrm{meV} \approx k_BT_M$.\cite{IOMN10u} We speculate that charge fluctuations\cite{Bale10} and accompanying spin fluctuations over this gap might be involved in the enhanced density of spin excitations. Topological Mott insulating states have been proposed for some of these systems\cite{RQHZ08,*PeBa10,*YaKi10}, but spin effects in a 3D topological insulator are confined to the sample surface and seem unlikely to contribute to the bulk muon spin relaxation. A spectroscopic study of low-lying fluctuations and $\Delta_g$ in Eu$_2$Ir$_2$O$_7$ would elucidate the situation, as would $\mu$SR experiments on the frustrated hyper\-kagom\'e iridate~Na$_4$Ir$_3$O$_8$\cite{ONA-KT07} and the (unfrustrated) weak Mott insulator~Sr$_2$IrO$_4$.\cite{KJMK08} 
 
In summary, the uniform spontaneous local field observed at muon sites below the MI/AFM transition indicates that Eu$_2$Ir$_2$O$_7$ exhibits long-range magnetic order, ruling out both quantum-spin-liquid (at least within the $\mu$SR time window) and spin-glass ground states. The magnetic structure cannot be obtained from $\mu$SR experiments alone, and neutron scattering in iridates is prohibitively difficult because of the high neutron absorption cross-sections of Ir nuclei. Resonant x-ray magnetic diffraction would be a useful alternative.

The dynamic muon spin relaxation rate~$\lambda_d(T)$ shows no sign of critical slowing down above $T_M$, suggesting a mean-field-like transition, and in the ordered state $\lambda_d(T)$ reveals an anomalous persistence of slow Ir$^{4+}$ spin fluctuations to low temperatures. Although geometric frustration may play a role in this behavior, the weakness of frustration in Eu$_2$Ir$_2$O$_7$, evidenced by the relatively large transition temperature, leads us to speculate that low-lying excitations associated with small-gap insulating behavior may be involved. Studies of other iridates, frustrated and unfrustrated, are clearly desirable.

\begin{acknowledgments}
We are grateful for technical assistance from the staff of the TRIUMF Center for Molecular and Materials Science. We would like to thank W.~P. Beyermann and J. Morales for their help with the susceptibility measurements, and L. Balents, S. Raghu, and C.~M. Varma for useful comments. This work was partially supported by the U.S. NSF, Grants 0801407 (UCR) and 0604105 (CSULA), by a Grant-in-Aid (No.~21684019) from JSPS, Japan, and by a Grant-in-Aid for Scientific Research on Priority Areas (19052003) from MEXT, Japan. 
\end{acknowledgments}

%\bibliography{abbrev,comments,critical,frust,iridates,kagome,math,Mott,muSR,NMR,%

\begin{thebibliography}{10}
\bibitem{quasistatic}%
  \BibitemOpen
  \bibinfo {note} {A muon local-field component~$\langle \mathbf{B}_{\rm
  loc}\rangle$ is quasistatic if it varies slowly compared to the muon spin
  precession period~$2\pi/\gamma_\mu\langle B_{\rm
  loc}\rangle$.~\protect\cite{HUIN79} We include the static limit in our use of
  this term. The dynamic muon spin relaxation time in zero or low applied field
  is then a lower bound on the correlation time of $\langle \mathbf{B}_{\rm
  loc}\rangle$.}%
  \bibAnnoteFile{Stop}{quasistatic}%
\end{thebibliography}
%pyrochlore,RE_compounds,skutt,spinglas,triangular}

%Merlin.mbs v4.21 2009-07-09.
%

\end{document}